\documentclass[11pt]{article}
\usepackage{amssymb}\usepackage{amsmath}
\newcommand{\fr}[2]{{\textstyle \frac{#1}{#2} }}

\def\la{\lambda}
\def\ot{\otimes}

\def\klein{\scriptscriptstyle}
\def\V{{\klein V}}
\def\HH{\mathbb H}

\def\CC{\mathbb C}
\def\EE{\mathbb E}

\def\CF{\mathcal F}
\def\CM{\mathcal M}
\def\bc{\big\langle \phi_{\{\la\}}\{z_i\}%
\big\rangle^{\Lambda^*\!,\infty}_{\Lambda,z_t}}
\def\bcr{\big\langle \phi_{\{\la\}}\{z_i\}%
\big\rangle^{\Lambda^*\!,r}_{\Lambda,z_t}}
\def\bcw{\big\langle \phi_{\{\la\}}\{w_i\}%
\big\rangle^{\Lambda^*\!,\infty}_{\Lambda,0}}
\def\bcy{\big\langle \{w_i\},y \big\rangle}

\begin{document}

\begin{center}
{\Large\bf
On SLE martingales in boundary WZW models
} \\ [4mm]
{\sc A.~Alekseev$^{1}$, A.~Bytsko$^{2,3}$, K.~Izyurov$^{1,3}$}  \\ [2mm]
{ \small
$^{1}$ Section of Mathematics, University of Geneva\\ 
 2--4 rue de Li\`evre, C.P. 64, 1211 Gen\`eve 4, Switzerland \\ [2mm]
$^{2}$  Steklov Mathematics Institute\\
 Fontanka 27, 191023, St.~Petersburg, Russia  \\ [2mm]
$^{3}$  Chebyshev Laboratory, St.~Petersburg State University\\
Universitetskaya nab. 7-9, 199034, St.~Petersburg, Russia 
} \\ [2.5mm]
{ }

\end{center}

\vspace{1mm}
\begin{abstract}
\noindent
Following \cite{BGLW}, we consider the boundary WZW model on a 
half--plane with a cut growing according to the Schramm--Loewner 
stochastic evolution and the boundary fields inserted at the tip 
of the cut and at infinity.
We study necessary and sufficient conditions for boundary correlation 
functions to be SLE martingales. Necessary conditions
come from the requirement for the boundary field at the tip of the cut 
to have a depth two null vector. Sufficient conditions 
are established using Knizhnik--Zamolodchikov equations 
for boundary correlators. 
Combining these two approaches, we show that in the case of $G=SU(2)$
the boundary correlator is an SLE  martingale 
if and only if the boundary field carries spin~$1/2$.
In the case of $G=SU(n)$ and $k=1$ there are several situations when
boundary one--point correlators are SLE$_\kappa$--martingales.
If the boundary field is labelled by the defining
$n$--dimensional representation of $SU(n)$, we obtain $\varkappa=2$.
For $n$ even, by choosing the boundary field labelled by the 
(unique) self--adjoint fundamental representation, we get
$\varkappa=8/(n {+} 2)$.
We also study the situation when the distance between the two 
boundary fields is finite, and we show that in this case 
the ${\rm SLE}_\varkappa$ evolution is replaced by  
${\rm SLE}_{\varkappa,\rho}$ with $\rho=\varkappa -6$.
 
\end{abstract}

\section*{Introduction}
  
Random conformally invariant curves often appear in
the scaling limit of interfaces in 2D statistical models
at critical points, see \cite{BB2,Ca1,Gr1,Sm1} for reviews. 
Such curves, if they have a Markov property, are
described by the Schramm--Loewner evolution (SLE).
Specifically, let a random conformally invariant Markov curve 
$\gamma_t$ start at the origin of the upper half plane~$\HH$. 
The parameter $t \geq 0$ can be regarded as the time of 
evolution. The seminal result of Schramm \cite{Sh1} states that
the dynamics of the tip $z_t$ of the curve 
is given by the law $z_t=g^{-1}_t(\sqrt{\varkappa} \xi_t)$, where 
$g_t(z)$ is the uniformizing conformal map which maps 
the slit domain $\HH\slash{\gamma_t}$ 
back to~$\HH$ and which satisfies the following 
stochastic differential equation:
\begin{equation}\label{dGt}
  dg_t(z) = \frac{2dt}{g_t(z) - \sqrt{\varkappa}\xi_t}  \,,
 \qquad g_0(z) = z \,.
\end{equation}
Here $\xi_t$ is the normalized Brownian process on $\mathbb R$,
starting at the origin, i.e., $\xi_0=0$, and $\EE[d\xi_t d\xi_t]=dt$.   
The parameter $\varkappa >0$ is the diffusion coefficient of the 
Brownian motion, and thus it is also an important parameter of the 
SLE trace.
 
The interplay between SLE and boundary conformal field theory has been 
studied in detail in the case of minimal models \cite{BB1} 
(see also \cite{BB2} for a review).
Consider the boundary minimal model ${\cal M}(p,p')$
($p$ and $p'$ are co--prime integers such that $p'>p\geq 2$) 
on the slit domain $\HH\slash{\gamma_t}$, where $\gamma_t$ is 
an SLE trace. 
Insert the boundary changing operators, $\phi$ and $\phi^\dagger$, 
at the tip $z_t$ of $\gamma_t$ and at $z=\infty$, respectively.
This insertion introduces two different boundary conditions, one on
the semi--axis from $-\infty$ to $z_t$, and the other one on the 
semi--axis from $z_t$ to $+\infty$.
Let $\cal O$ stand for a set of primary operators at fixed points 
in the bulk. It was observed in \cite{BB1} that the normalized 
boundary correlation function
\begin{equation}\label{Ocf}
 {\cal M}_t=
 \frac{\langle \phi(z_t) {\cal O} \phi^\dagger(\infty) \rangle}%
 {\langle \phi(z_t) \phi^\dagger(\infty)\rangle}  
\end{equation}
is an SLE  martingale. That is, it is conserved in mean under  
${\rm SLE}_\varkappa$, 
$ \EE[\frac{d}{dt} {\cal M}_t]=0$, provided 
that $\varkappa=4p'/p$ or $\varkappa=4p/p'$ and $\phi$ is the 
primary operator $\phi^{p,p'}_{1,2}$ or
$\phi^{p,p'}_{2,1}$, respectively. 

Since analytic properties of CFT correlation functions 
are well understood (see, e.g. \cite{DMS}), existence of 
martingales of type (\ref{Ocf}) can be exploited in 
computation of various SLE related probabilities, 
see e.g.~\cite{BB2}. This is a motivation to search 
for new martingales in non--minimal boundary CFTs. 
For the $SU(2)$ WZW model,
some results in this direction were obtained in \cite{BGLW,Ra1}. 
The aim of this paper is to better understand and extend the results 
of~\cite{BGLW}.

The paper is organized as follows. In Section~1, we show that
a boundary correlation function of the WZW model with a boundary 
field $\phi_\Lambda$ inserted at the tip of an SLE
trace is an SLE$_\varkappa$ martingale if a certain descendant of 
$\phi_\Lambda$ is a level two null vector with respect to 
the Kac--Moody algebra~$\hat{\mathfrak{g}}_k$. In comparison
to the minimal models, one has 
to assume in addition that the evolution of the SLE trace is accompanied
with a random gauge transformation of the bulk fields \cite{BGLW}. 
The randomness of the gauge transformation is described
by a Brownian motion on the group with a  coupling constant $\tau$.
This is an additional parameter which must be adjusted to the value 
of~$\varkappa$.

In Section~2, we analyse necessary conditions for
the null vector ensuring the martingale property of the correlation 
function.
We show that, for a given Lie algebra ${\mathfrak g}$, these 
conditions are satisfied for more than two different values 
of $k$ (and thus there can be more than two different 
values of $\varkappa$) only if $\text{dim}\,{\mathfrak g}=3$. 
Furthermore, for $\mathfrak{g} = \mathfrak{su}(2)$, we show that
$\Lambda$ must be the fundamental representation (i.e.
corresponding to spin~$1/2$), and $\varkappa=\frac{4(k+2)}{k+3}$
unless $k=1$ (if $k=1$, $\varkappa$ is not fixed).
This confirms the conclusions of~\cite{BGLW}.
For $\mathfrak{g} = \mathfrak{su}(n)$ with $n>2$, we show that
when $\Lambda$ is the fundamental representation,
the necessary conditions imply $k=1$ and $\varkappa=2$. 
For non--fundamental representations $\Lambda$ and $k=1$, 
the necessary conditions imply $\varkappa=\frac{8}{n+2}$
provided that the Casimir operator $C_\Lambda$ acquires
a certain value. We show that this condition holds for
all even $n$ for a self--conjugate $\Lambda$ of a specific form.

In Section~3, we use the Knizhnik--Zamolodchikov equations
to derive a sufficient condition ensuring the martingale property.
More precisely, we show that the correlation function is a martingale
if it is contained in the kernel of a certain matrix. 
For $\mathfrak{g} = \mathfrak{su}(2)$ and  $k>1$, we observe that 
under the necessary conditions of Section~2 
the matrix in question vanishes, and the necessary conditions turn
out to be sufficient.
 
In Section~4, we consider explicit expressions
for boundary correlation functions with one bulk field. We study the 
situation when the $\mathfrak{g}$--invariant submodule is
two--dimensional but the corresponding space of conformal blocks is 
one--dimensional due to the fusion rules at the level $k=1$.
We show that, for the weights $\Lambda$ allowed by the
necessary conditions and the corresponding values of $\varkappa$
found in Section~2, the one--point boundary correlators
are indeed SLE$_\kappa$ martingales.

In Section~5, we consider the case when the second boundary
operator is inserted at a finite distance from the origin.
We show that the corresponding boundary correlator is
an ${\rm SLE}_{\varkappa,\rho}$ martingale if $\rho=\varkappa-6$
and the  null vector condition of Section~2 holds. 
 We use the KZ equation to derive 
a sufficient condition similar to that found in Section~3.

\pagebreak[2] 
\section{SLE martingales in WZW}
 
Let $\mathfrak{g}$ be a simple Lie algebra.
We study the boundary 
$\hat{\mathfrak{g}}_k$  WZW model on 
the slit domain $\HH\slash{\gamma_t}$, where 
$\gamma_t$ is an $\text{SLE}_\varkappa$ trace. 
Consider a boundary correlation function  with $N$ primary fields 
in the bulk,
where the field $\phi_{\la_i}(z_i)$ ($i=1,{\ldots},N$, $\Im(z_i)>0$) 
has a  conformal weight $h_i$ and carries an irreducible 
$\mathfrak{g}$ representation of a highest weight~$\la_i$.
The boundary condition changing operators, 
$\phi_{\Lambda}$ and $\phi_{\Lambda^*}$, are 
inserted at the tip $z_t$ of $\gamma_t$ and at $z=\infty$. 
The boundary correlation function \cite{Ca2} for this set of fields 
is a certain chiral conformal block (the choice of a particular 
conformal block
depends on the boundary conditions) for the theory on the complex 
plane $\CC$
with additional primary fields corresponding to
conjugate  representations $\lambda_i^*$ placed 
at the mirror image points $\bar{z}_i$,
\begin{equation}\label{bound}
\begin{aligned}
 \bc {}& \equiv  \frac{
 \langle \phi_{\Lambda}(z_t)
  \phi_{\la_1}(z_1) \ldots \phi_{\la_N}(z_N) \,
  \phi_{\la_1^*}(\bar{z}_1) \ldots
    \phi_{\la_N^*}(\bar{z}_N) \phi_{\Lambda^*}(\infty) \rangle^\mathfrak{g}
 }{ \langle \phi_{\Lambda}(z_t) \phi_{\Lambda^*}(\infty) 
 \rangle^\mathfrak{g} } \,.
\end{aligned}
\end{equation}
Here the numerator takes values in the $\mathfrak{g}$--invariant
subspace of the tensor product 
$V_\Lambda \,{\ot}\, V_{\lambda_1} \,{\ot}\, {\ldots}
 \,{\ot}\, V_{\Lambda^*}$. 
The denominator takes values in the $\mathfrak{g}$--invariant
subspace of $V_\Lambda \,{\ot}\, V_{\Lambda^*}$, which by 
the Schur's lemma 
is one--dimensional, and so the denominator is a scalar. 
The $\mathfrak{g}$--invariance of the correlation function 
is expressed by the equation,
\begin{equation}\label{ta}
  \Bigl(\sum_{i=0}^{2N+1} t^a_i \Bigr) \bc =0 ,
\end{equation}
where $t^a$ form an orthonormal basis of $\mathfrak{g}$, and $t^a_i$ is 
a matrix representing $t^a$ in the $i$--th tensor factor
in the representation of a highest weight $\lambda_i$, $t^a_0$ acts on 
$\phi_\Lambda(z_t)$ and $t^a_{2N+1}$ acts on $\phi_{\Lambda^*}(\infty)$.

It is convenient to introduce the conformal map 
$w_t(z) = g_t(z) - \sqrt{\varkappa}\xi_t$.
The dynamics of the tip of the SLE trace is then given 
by $z_t=w^{-1}_t(0)$.
The map $w_t(z)$ satisfies the 
stochastic differential equation:
\begin{equation}\label{dz}
    dw_t(z) = \frac{2 dt}{w_t(z) }   - \sqrt{\varkappa} d\xi_t 
\end{equation}
with initial condition $w_0(z)=z$.
The map $w_t(z)$ maps the initial configuration of fields on
the slit domain $\HH\slash\gamma_t$ into a configuration on the upper 
half plane~$\HH$. The boundary condition changing operators 
$\phi_\Lambda$ and $\phi_{\Lambda^*}$ are now inserted at $w=0$ and 
$w=\infty$, and the bulk primary fields $\phi_{\lambda_i}$ are
positioned at the points $w_i \equiv w_t(z_i)$ 
which are moving as $t$ increases. For the theory on $\HH$,
it is well known \cite{Ca2} that the mirror images of bulk fields  
are located at the complex conjugate points, that is 
$w_{i+N}=\overline{w_i}$, $i=1,{\ldots},N$.
Note that solutions of equation (\ref{dz}) satisfy the reflection
property, $\overline{w_t(z)}=w_t(\overline{z})$.
Therefore, in (\ref{bound}), we also have pairs of conjugate points
$z_i, \bar{z}_i$.

Since $w_t(z)$ is a conformal map, we can rewrite 
the correlation function (\ref{bound})
in the new coordinates: 
\begin{align}
\label{bc2a}
  \bc 
{}& = \biggl( \prod_{i=1}^{2N}
 \Bigl( \frac{\partial w_i}{\partial z_i} \Bigr)^{h_{i}}
 \!\biggr) \bcw ,
\end{align}
where $w_{i+N}=\bar{w}_i, z_{i+N}=\bar{z}_i$.

Let us determine the increment of (\ref{bc2a}) when $t$ is 
increased by~$dt$. Eq. (\ref{dz}) implies that the prefactor 
changes as follows: 
\begin{equation}\label{dwh}
\frac{d}{dt}\Bigl( \frac{\partial w_i}{\partial z_i} \Bigr)^{h}=
 h\Bigl( \frac{\partial w_i}{\partial z_i} \Bigr)^{h-1}
\partial_t \frac{\partial w_i}{\partial z_i}
= h\Bigl( \frac{\partial w_i}{\partial z_i} \Bigr)^{h-1}
 \partial_{z_i} \Bigl( \frac{2}{w_i}\Bigr)=
 - \frac{2h}{w_i^2}
 \Bigl( \frac{\partial w_i}{\partial z_i} \Bigr)^{h} \,.
\end{equation}
If we were considering a minimal model, the increment of a bulk 
field would have been given by
\begin{equation}\label{df}
 d\phi_{\la_i}(w_i) = {\cal G}_i \phi_{\la_i}(w_i) \,, 
\end{equation}
where, by (\ref{dz}), we would have had
${\cal G}_i = dw_i \partial_{w_i} = 
(\frac{2dt}{w_i} - \sqrt{\varkappa} d\xi_t) \partial_{w_i}$.
In the case of the WZW model, the fields are Lie group valued, and
one can introduce an additional random motion in 
the target space. 
The following modification was proposed in \cite{BGLW}: 
\begin{equation}\label{dP}
 {\cal G}_i  =
 \Bigl( \frac{2dt}{w_i} - \sqrt{\varkappa} d\xi_t \Bigr) \partial_{w_i} 
 + \frac{\sqrt{\tau}}{w_i} \sum_{a=1}^{\text{dim}\,{\mathfrak g}} 
 \Bigl(  d\theta^a \, t^a_i \Bigr)   \,, 
\end{equation}
where $d\theta^a$ are normalized generators of 
a ${\mathbb R}^{\text{dim}\,{\mathfrak g}}$--valued Brownian motion, 
i.e.,
\begin{equation}\label{EE}
 \EE (d\theta^a  d\theta^b) = \delta_{ab} \, dt    \,.
\end{equation}

Note that a Brownian motion on a Lie group $G$ is defined by the following
stochastic differential equation,
\begin{equation} \label{BrownG}
dg= \Bigl( \alpha \sum_a \, d\theta^a t^a + 
\frac{\alpha^2}{2} \sum_a t^at^a \, dt \Bigr) g.
\end{equation} 
Here the second term on the right hand side is taking care 
of the exponential map between
the Lie algebra and the Lie group. For instance, in the case 
of $G=U(1)$, we have
$g=\exp(i \alpha \xi_t)$, where $\xi_t$ is the one-dimensional 
Brownian motion.
Then, using the standard Ito calculus we obtain 
$$
dg= \Bigl( i \alpha d\xi_t - \frac{\alpha^2}{2} \, dt \Bigr) g.
$$
Equation (\ref{BrownG}) suggests the following alternative 
writing of equation 
(\ref{dP}):
\begin{equation}\label{dPnew}
 {\cal G}_i  =
 dt \, \Bigl( \frac{2}{w_i} \, \partial_{w_i} - 
   \frac{\tau C_i}{2w_i^2}\Bigr)
 - \sqrt{\varkappa} d\xi_t \partial_{w_i} 
 + \Bigl( \frac{\sqrt{\tau}}{w_i} \sum_{a} \, d\theta^a \, t^a_i
+ \frac{\tau}{2 w_i^2} \sum_a \, t^a_i t^a_i \, dt \Bigr)  \,, 
\end{equation}
where $C_i$ is the value of the quadratic Casimir operator
$\sum_a t^a_i t^a_i$ is the representation with highest weight 
$\lambda_i$. In operator (\ref{dPnew}), the first two terms correspond 
to the SLE developing on the upper half-plane, and the third term
describes the Brownian motion on the group.

Returning to the analysis of the boundary correlation function, 
let us introduce the following operator:
\begin{equation}\label{theta1}
 \Theta = \sum_{i=1}^{2N} \Bigl(
 \frac{2}{w_i} \, \partial_{w_i}
 -   \frac{2h_i}{w_i^2} \Bigl)
 + \frac{\varkappa}{2}   \sum_{i,j=1}^{2N}
 \partial_{w_i} \partial_{w_j}  
+ \frac{\tau}{2}
  \sum_{i,j=1}^{2N} \frac{T_{ij}}{w_i w_j}  \,,
\end{equation}
where $T_{ij} = T_{ji} \equiv \sum_a t^a_i \, t^a_j$. 
Let us show that the correlator $\bc$ is an SLE$_\varkappa$ 
martingale if and only if its $w$--image is annihilated by $\Theta$,
\begin{equation}\label{mart1}
 \Theta \, \bcw = 0 \,.
\end{equation}
Indeed, substituting (\ref{dP}) in (\ref{df}), using the Ito formula,
and taking into account (\ref{bc2a}) and (\ref{dwh}), we find  
\begin{align}
\nonumber
{}&
\biggl( \prod_{i=1}^{2N}
\Bigl( \frac{\partial w_i}{\partial z_i} \Bigr)^{-h_{i}} \!\biggr)
\EE\bigl[d \bc \bigr] \\
\nonumber
{}&=
  -\sum_{i=1}^{2N}  \frac{2h_i dt}{w_i^2} \, \bc + 
 \EE\bigl[d \bcw \bigr] \\
\nonumber
{}& =
 \Bigl( -\sum_{i=1}^{2N}  \frac{2h_i dt}{w_i^2} +
 \EE\Bigl[ \sum_{i=1}^{2N} {\cal G}_i +
 \fr{1}{2} \sum_{i,j=1}^{2N} {\cal G}_i {\cal G}_j \Bigr] \Bigr) \bcw \\
\label{boundGG1}
{}& = dt \, \Theta \, \bcw  .
\end{align} 

Recall (see, e.g., \cite{DMS}) that, if $X=\prod_i \phi_i(w_i)$, then
$\langle (L_{-n} \phi)(z) X \rangle =
 {\cal L}_{-n} \langle \phi(z) X \rangle$ for
$n \geq 1$ and $\langle (J^a_{-n} \phi)(z) X \rangle =
 {\cal J}^a_{-n} \langle \phi(z) X \rangle$ for $n\geq 0$,
where
\begin{align}
\label{CL}
 {\cal L}_{-n} = \sum_i \Bigl( \frac{(n-1)h_i}{(w_i-z)^n}
 - \frac{1}{(w_i-z)^{n-1}} \partial_{w_i} \Bigr) \,, \qquad
 {\cal J}^a_{-n} = - \sum_i \frac{t^a_i}{(w_i-z)^n} \,.
\end{align}
Therefore, the martingale condition (\ref{mart1}) can be rewritten 
as follows:
\begin{align}
\nonumber
 0 {}& = 
  \Bigl(
 -2{\cal L}_{-2} + \fr{1}{2} \varkappa  {\cal L}_{-1}^2
+ \fr{1}{2}\tau \sum_{a}   {\cal J}^a_{-1} {\cal J}^a_{-1}   
\Bigr) \bcw \\
\label{boundGG3}
{}& = \frac{
 \langle \psi(0) 
  \phi_{\la_1}(w_1) \ldots  
    \phi_{\la_N^*}(w_{2N}) \phi_{\Lambda^*}(\infty) \rangle^{\mathfrak g}
 }{ \langle \phi_{\Lambda}(0) \phi_{\Lambda^*}(\infty) 
 \rangle^{\mathfrak g} } \,,
\end{align}
where
\begin{equation}\label{psi}
 \psi = \bigl( -2 L_{-2} + \fr{1}{2} \varkappa L^2 _{-1}
 +  \fr{1}{2} \tau \sum_{a=1}^{\dim \mathfrak g} J_{-1}^a J_{-1}^a  
 \bigr) \phi_\Lambda  .
\end{equation}
Thus, a sufficient condition for the correlation function in 
question to be a covariant $\text{SLE}_\varkappa$ martingale
is the requirement that $\psi$ be a level two null vector.

\section{Null vectors and necessary conditions}

In this Section, we analyse in detail the null vector property 
of $\psi$ defined by  (\ref{psi}).
It is equivalent to two
equations, $J^a_1 \psi =0$ and $J^a_2 \psi = 0$.
Recall that the Kac--Moody and Virasoro generators
satisfy the following commutation relations:
\begin{align}
\label{LL}
 [L_m, L_{m'} ] &= (m-m') L_{m+m'} +
    \fr{c}{12} m(m^2-1) \delta_{m+m',0} \,,\\
\label{LJ}
 [L_m, J^a_{m'} ] &= -m' J^a_{m+m'} \,, \\
\label{JJ}
 [ J^a_m, J^b_{m'} ] &= \sum_c i f_{abc} J^c_{m+m'} +
    k m \delta_{ab} \delta_{m+m',0} \,,
\end{align}
where $k$ is the level and $c$ is the central charge given by
\begin{equation}\label{ck}
 c= \frac{k \text{\,dim\,} {\mathfrak g}}{k + h^\V} \,.
\end{equation}
Here $h^\V$ is the dual Coxeter number of the Lie algebra $\mathfrak{g}$.

Acting with $J_1^b$, and $J_2^b$ on (\ref{psi}) we obtain
\begin{align}
 \label{jp1}
{}& \Bigl(  \bigl(\tau k - \tau h^\V  -2\bigr) J^b_{-1}
 + \varkappa J^b_0 L_{-1} + i \tau \sum_{a,c} f_{abc} J^a_0 J^c_{-1}
 \Bigr) \phi_{\Lambda} = 0 \,,\\
\label{jp2}
{}& \bigl( \varkappa + \tau h^\V -4
 \bigr) J^b_0 \phi_{\Lambda} = 0 \,.
\end{align}
Here we used that
$\sum_{a,c} f_{bac} f_{dac} = 2 h^\V \delta_{bd}$.
Equations (\ref{jp1})--(\ref{jp2}) define a necessary and 
sufficient conditions for $\psi$ given by (\ref{psi}) to 
be a null vector. Equation (\ref{jp2})
 implies that (here we assume $\Lambda \neq 0$)
\begin{equation}\label{taukap}
 \varkappa + \tau h^\V = 4 \,.
\end{equation}
Equation (\ref{jp1}) is more involved. However, acting on 
it with $L_1$, we derive the following (simpler) necessary condition: 
\begin{equation}\label{jp3}
\bigl( 2 \varkappa h_\Lambda + \tau k  -2 \bigr)
 J^b_0 \phi_{\Lambda} = 0 \,.
\end{equation}
Another  necessary condition can be obtained
by requiring $L_2 \psi=0$, which yields
\begin{equation}\label{jp4}
 \bigl( 3 \varkappa h_\Lambda + \fr{1}{2} \tau c (k + h^\V)
 -8 h_\Lambda -c \bigr) \phi_{\Lambda} = 0 \,.
 \end{equation}
In (\ref{jp3}) and (\ref{jp4}), it was used that 
$L_0 \phi_\Lambda=h_\Lambda\phi_\Lambda$. Recall
that the conformal dimension $h_\Lambda$ is given by
\begin{equation}\label{hlam}
 h_{\Lambda} = \frac{ C_\Lambda}{2(k+h^\V) } \,,
\end{equation}
where
$C_\Lambda$ 
is the value of the Casimir operator $C=\sum_a t^a t^a$ in 
the irreducible representation of a highest
weight $\Lambda$.

For $k \neq 2h_\Lambda h^\V$, eqs. (\ref{taukap})--(\ref{jp3})
imply that
\begin{equation}\label{oth4}
   \varkappa = \frac{2(h^\V -2k)}{2h_\Lambda h^\V -k} \,, \qquad
   \tau = \frac{8 h_\Lambda -2 }%
        {2h_\Lambda h^\V -k} \,.
\end{equation}
Substituting (\ref{ck}) and (\ref{hlam})--(\ref{oth4})
in (\ref{jp4}), we arrive at the following condition
\begin{equation}\label{ch0}
\begin{aligned}
{}& \bigl( h^\V \text{\,dim\,} {\mathfrak g} +
 2C_\Lambda(1-\text{\,dim\,} {\mathfrak g}) \bigr) k^2 \\
{}& \qquad\qquad + 
 \bigl( h^\V \text{\,dim\,} {\mathfrak g} -
 C_\Lambda(1+\text{\,dim\,} {\mathfrak g})  \bigr) h^\V k
 + 4 C_\Lambda^2 h^\V - 3 C_\Lambda (h^\V)^2 =0 \,.
\end{aligned}
\end{equation} 

Let us analyse eqs. (\ref{taukap})--(\ref{jp4}) and 
(\ref{ch0}) in some particular cases.

0) For $k = 2h_\Lambda h^\V$, formulae (\ref{oth4}) do not apply.
In this case, eqs. (\ref{taukap}) and (\ref{jp3}) are linearly 
dependent and they have a solution only if 
\begin{equation}\label{khcexc}
   k= \fr{1}{2} h^\V  \,, \qquad  h_\Lambda = \fr{1}{4} \,, 
	\qquad C_\Lambda = \fr{3}{4} h^\V \,.
\end{equation}
Note that the condition $C_\Lambda = \frac{3}{4} h^\V$ cannot
hold for ${\mathfrak g}=\mathfrak{su}(n)$, $n>2$.
Under conditions (\ref{khcexc}), eq. (\ref{jp4}) is equivalent to
\begin{equation}\label{kapdimexc}
 (3 \varkappa -8)(\text{\,dim\,} {\mathfrak g} -3)=0 \,.
\end{equation}
For ${\mathfrak g}=\mathfrak{su}(2)$, this relation holds for 
any $\varkappa$, and the condition $C_\Lambda = \frac{3}{4} h^\V$ 
implies that $\Lambda$ is the representation of spin~$1/2$. 
Thus, the case ${\mathfrak g}=\mathfrak{su}(2)$, $k=1$,
$\Lambda$ being the fundamental representation
is very degenerate, the parameter $\varkappa$ is not fixed and
the only  relation imposed on $\varkappa$ 
and $\tau$ is eq.~(\ref{taukap}).

1) ${\mathfrak g}=\mathfrak{su}(2)$, $h^\V=2$. 
Let $\Lambda$ be a representation of spin~$j$. We have
$h_\Lambda=j(j+1)/(k+2)$, and then (\ref{ch0}) is equivalent 
to the condition
$(2j-1) (2j+3) (2j-k) (k+2j+2)=0$. That is, 
either $j=1/2$ or $k=2j$. The latter possibility is actually
excluded since it corresponds to the case of $k = 2h_\Lambda h^\V$.
Thus, $\Lambda$ must be the fundamental representation, i.e.
of spin~$1/2$. Then, for $k\neq 1$, eqs. (\ref{oth4}) yield
\begin{equation}\label{oth5}
   \varkappa = \frac{4(k+2)}{k+3} \,, \qquad
   \tau = \frac{2}{k+3} \,.
\end{equation}
 
Note that ${\mathfrak g}=\mathfrak{su}(2)$ is the 
only case, when, for a given $C_\Lambda$, condition 
(\ref{ch0}) can hold for more than two different values of $k$ 
(and thus for any $k$). Indeed, the polynomial in $k$ 
given by the l.h.s. of (\ref{ch0}) is identically zero
only if
\begin{equation}\label{kinf}
    h^\V \text{\,dim\,} {\mathfrak g} =
      2 C_\Lambda \, (\text{\,dim\,} {\mathfrak g} -1) \,,
 \qquad \text{and} \qquad \text{dim}\, {\mathfrak g} = 3 \,.
\end{equation}
For a simple Lie algebra, the second condition implies that 
${\mathfrak g}=\mathfrak{su}(2)$.
Then, the first condition implies that 
$\Lambda$ is the representation of spin $j=1/2$.  
 
2a) ${\mathfrak g}=\mathfrak{su}(n)$, $h^\V=n>2$. 
Let $\Lambda$ be the fundamental representation. 
We have $C_\Lambda=(n^2-1)/n$, and
(\ref{ch0}) is equivalent to the condition
$(k^2 -1)(n^2-1)(n^2-4)=0$. Whence, for
$n>2$, the only possibility is $k=1$. 
In this case, (\ref{oth4}) yields
\begin{equation}\label{oth6}
   \varkappa = 2 \,, \qquad
   \tau = \frac{2}{n} \,.
\end{equation}

2b) ${\mathfrak g}=\mathfrak{su}(n)$, $h^\V=n>2$. 
Consider the case of $k=1$.  Then,
(\ref{ch0}) is satisfied either if
$C_\Lambda=(n^2-1)/n$ (and we recover the case 2a),
or if
\begin{equation}\label{specc}
   C_\Lambda=\frac{n(n+1)}{4} \,.
\end{equation} 
This condition holds
for self--conjugate representations whose 
Dynkin labels are
$\Lambda \sim (0,1,0)$, $\Lambda \sim (0,0,1,0,0)$, etc.
Here $n$ is required to be even.
In this case, we have $h_\Lambda = n/8$, and
(\ref{oth4}) yields
\begin{equation}\label{oth7}
   \varkappa = \frac{8}{n+2} \,, \qquad
   \tau = \frac{4}{n+2} \,.
\end{equation} 
It is interesting that the set of values of $\varkappa$ in 
equation (\ref{oth7})
does not meet the set of values of $\varkappa$ in equation (\ref{oth5}).
Moreover, $\varkappa$'s of equation (\ref{oth7})
are  not contained in the set 
corresponding to the minimal model $M(p,p')$.
Indeed, (\ref{oth7}) matches $\varkappa=4p/p'$ for $p=2$ and 
$p'=n+2$, but $p'$ must be co--prime with~$p$, 
which is not the case when $n$ is even.
 
2c) ${\mathfrak g}=\mathfrak{su}(n)$, $h^\V=n>2$. 
If $\Lambda$ is the adjoint representation,
$\Lambda \sim (1,0,{\ldots},0,1)$, then
$C_\Lambda=2n$. In this case,
(\ref{ch0}) is equivalent to the condition
$(3n^2-7)k^2 +n(n^2+1)k -10n^2=0$. For $n>2$,
the only positive integer solution is
$n=7$, $k=1$. However, for $k=1$
the adjoint representation does not satisfy the 
integrability constraint, $|\Lambda| \leq k=1$.

\section{KZ equations and sufficient conditions}

In this Section we obtain and study sufficient conditions 
for a correlation function to be an SLE martingale.

Below we will use the notation $\nu \equiv 1/(k\,{+}\,h^\V)$.
Recall that $T_{ij} \equiv \sum_a t^a_i t^a_j$.
Using that  $T_{ii} \phi_{\lambda_i} = (2h_{i}/\nu) \phi_{\lambda_i}$,
we can rewrite (\ref{boundGG1}) as follows: 
\begin{equation}\label{Et2}
 \Theta  =  
\biggl( \sum_{i=1}^{2N} \Bigl(
 \frac{2}{w_i} \, \partial_{w_i}
 -   \frac{2H_i}{w_i^2} \Bigl)
 + \frac{\varkappa}{2}\!   \sum_{i,j=1}^{2N}
 \partial_{w_i} \partial_{w_j}  
+ \tau\!
  \sum_{i<j}^{2N} \frac{T_{ij}}{w_i w_j}   
 \biggr) \,
\end{equation}
where  
\begin{equation}\label{Hi}
  H_i = h_i (1 - \frac{\tau}{2\nu} ) 
\end{equation}
are renormalized conformal weights. Note that the renormalization
of conformal weights is similar to the redistribution of terms
in the operator ${\cal G}_i$ in equation (\ref{dPnew}).

Correlation functions of the WZW model satisfy the 
Knizhnik--Zamolodchikov (KZ) equations~\cite{KZ}.
In our case, they read
\begin{equation}\label{kz2}
  {\partial_{w_i}} \bcw
 = \nu \biggl( \frac{T_{0i}}{w_{i}} +
  \sum_{j\neq i}^{2N} \frac{T_{ij}}{w_{i}-w_{j}} \biggr) \,
   \bcw .
\end{equation}
Hence, we have
\begin{align} 
\label{kzNN}
{}& 
  \sum_{i,j=1}^{2N}  {\partial_{w_i}}{\partial_{w_j}} \bcw 
  = \biggl( \nu ^2\!
 \sum_{i,j=1}^{2N} \frac{T_{0i} T_{0j}}{w_{i}w_{j}}
  - \nu \sum_{i=1}^{2N} \frac{T_{0i}}{w_{i}^2} \biggr) \, \bcw,\\
\label{kzNNN}
{}& \sum_{i=1}^{2N}  \frac{1}{w_{i}} {\partial_{w_i}} \bcw
 = \nu \biggl( \sum_{i=1}^{2N}  \frac{T_{0i}}{w_{i}^2} -
 \sum_{ i<j}^{2N} \frac{T_{ij}}{w_{i} w_{j}} \biggr) \bcw .
\end{align}  
Applying the operator (\ref{Et2}) to the correlation function
and using these identities, we rewrite the martingale condition
(\ref{mart1}) as an algebraic equation: 
\begin{equation}\label{dPH2} 
 \CM\bigl(\{w_{i}\}\bigr) \, \bcw =0 \,,  
\end{equation}
where
\begin{align} 
\label{Mw}
{}& \CM\bigl(\{w_{i}\}\bigr) =
 \sum_{i=1}^{2N} \frac{A_i}{w_{i}^2} +
 \sum_{i<j}^{2N} \frac{B_{ij}}{w_{i} w_{j}} ,\\
 \label{Aj}
{}&  A_i = (4 -\varkappa) \, T_{0i} + \varkappa\nu \, \bigl(T_{0i}\bigr)^2
   + \fr{1}{\nu} h_i (\fr{2}{\nu} \tau -4) 
 \,,   \\[1mm]
\label{Bj}
{}&  B_{ij} = (\fr{2}{\nu} \tau -4) \, T_{ij} + \varkappa\nu \,
  \bigl( T_{0i} T_{0j} + T_{0j} T_{0i} \bigr) \,.
\end{align}
Thus, the boundary correlation function
in question is a martingale if
it is in the kernel of the matrix~$\CM\bigl(\{w_{i}\}\bigr)$.

Recall that, for $\mathfrak{g}=\mathfrak{su}(2)$,
the necessary conditions require $\Lambda$ to be the 
representation of spin~$1/2$ and, for $k\neq 1$, 
\begin{equation}\label{kth2}
  \varkappa = \frac{4}{\nu +1} \,, \qquad
  \tau = \frac{2\nu}{\nu +1} \,,  
\end{equation}
where $\nu=1/(k+2)$.
The properly normalized generators of $\Lambda$ are 
$t^a_0 = \fr{1}{\sqrt{2}} \sigma^a$, where $\sigma^a$
are the Pauli matrices. Using their properties, 
we readily derive the following identities:
\begin{align}
 \label{T2a}
{}& \bigl(T_{0i}\bigr)^2 = \fr{1}{2} (\sigma^a t^a_i)
 (\sigma^b t^b_i) =
  \fr{1}{2} (t^a_i t^a_i  - \sqrt{2}\sigma^a t^a_i) =
 \fr{1}{\nu} h_i - T_{0i}  \,, \\
 \label{T2ab}
{}& T_{0i} T_{0j} + T_{0j} T_{0i} =
 \fr{1}{2} (\sigma^a \sigma^b + \sigma^b \sigma^a) \,
  t^a_i  t^b_j = T_{ij} \,.
\end{align}
Substituting (\ref{kth2})--(\ref{T2ab}) in (\ref{Aj}) 
and (\ref{Bj}), we obtain
\begin{equation}\label{AB0}
 A_i =0 \,, \qquad B_{ij} = 0 \,.
\end{equation}
Hence, in this case, $\CM\bigl(\{w_{i}\}\bigr)$ 
vanishes identically. 
In conclusion, we have proved that for 
$\mathfrak{g}=\mathfrak{su}(2)$  and $k\neq 1$ a boundary 
correlation function is an 
SLE$_\varkappa$ martingale if and only if the boundary field 
is in the fundamental representation, and $\varkappa$ and $\tau$
are given by (\ref{kth2}).
We will show below that in the special case of $k=1$ it
is possible that $\CM\bigl(\{w_{i}\}\bigr)$ does not vanish but
has a non--empty kernel and a certain conformal block lies in the kernel.

Relations (\ref{AB0}) are sufficient but not necessary conditions
for the martingale property. In fact, their weaker form,
$A_i^{\mathfrak g} =0$ and $B_{ij}^{\mathfrak g} = 0$,
is also a sufficient condition.
Here the superscript ${\mathfrak g}$ denotes the projection
onto the ${\mathfrak g}$--invariant subspace of
$V_\Lambda \,{\ot}\, V_{\lambda_1} \,{\ot}\, {\ldots}
 \,{\ot}\, V_{\Lambda^*}$.
However, even this form produces a very restrictive condition.
In particular, $A_i^{\mathfrak g} =0$ implies that
$T_{0i}^{\mathfrak g}$ has at most two distinct eigenvalues.
In the $\mathfrak{su}(2)$ case, this is true since
$V_{1/2} \otimes V_{j} = V_{j-1/2} \oplus V_{j+1/2}$.
 
Consider the case ${\mathfrak g} =\mathfrak{su}(n)$, $n>2$,
$\Lambda$ being the fundamental representation.
If $\la_i =\Lambda$ or $\la_i =\Lambda^*$, then
$T_{0i}$ has exactly two distinct eigenvalues 
(cf.~Section~4.1):
$\bigl(\frac{n-1}{n},\frac{-n-1}{n}\bigr)$ and
$\bigl(\frac{1}{n},\frac{1-n^2}{n}\bigr)$, respectively.
On the other hand, for $\Lambda$ being the fundamental representation 
in the $n>2$ case, the necessary  conditions imply $k=1$,
and $\varkappa$ and $\tau$ must be given by equation (\ref{oth6}).
For these data, equation $A_i^{\mathfrak g} =0$ implies that
the eigenvalues of $T_{0i}^{\mathfrak g}$ are equal to 
$\bigl(\frac{1-n^2}{n},\frac{-n-1}{n}\bigr)$.
Thus,
$\CM\bigl(\{w_{i}\}\bigr)$ does not vanish. 
However, similarly to the $\mathfrak{su}(2)$ case, we will show 
that $\CM\bigl(\{w_{i}\}\bigr)$ may have a non--empty kernel 
containing a certain conformal block.

\section{One--point boundary correlators for  $k=1$}

In this Section, we consider explicit expressions for boundary
one--point correlation functions (that is, the case of  $N=1$).
The normalized correlator is of the form,
\begin{equation}\label{N1}  
 \big\langle \phi_{\la}(w)%
\big\rangle^{\Lambda^*\!,\infty}_{\Lambda,0} \equiv 
 \frac{ \langle \phi_{\Lambda}(0) \phi_{\la}(w)  
  \phi_{\la^*}(\bar{w}) \phi_{\Lambda^*}(\infty) \rangle^{\mathfrak g}
 }{ \langle \phi_{\Lambda}(0) \phi_{\Lambda^*}(\infty) 
\rangle^{\mathfrak g} } \,.
\end{equation}
Recall that the $SL(2,\CC)$ invariance implies that
(see, e.g.~\cite{DMS})
\begin{equation}\label{bound2}
 \big\langle \phi_{\la}(w)%
\big\rangle^{\Lambda^*\!,\infty}_{\Lambda,0}  =
 (\bar{w})^{-2 h_{\la}} \, \CF(x)  \,, \qquad x = w/\bar{w} \,,
\end{equation}
where $\CF(x)$ belongs to the
$\mathfrak{g}$--invariant submodule $W^{\mathfrak{g}}$ of
$V_\Lambda \,{\ot}\, V_{\lambda} \,{\ot}\, V_{\lambda^*}
 \,{\ot}\, V_{\Lambda^*}$.
Substituting this expression in the KZ equations (\ref{kz2})
(the variables $w_1=w$ and $w_2=\bar{w}$ are regarded as 
independent) yields 
\begin{align}
\label{kz3a}
  \Bigl( \fr{1}{\nu} \partial_{x} -
  \frac{T_{01}}{x} - \frac{T_{12}}{x-1} \Bigr) \,
  \CF(x)  = 0 \,,\\ 
  \label{kz3c}
  \Bigl( \fr{1}{\nu} x \partial_{x} + \fr{2h_{\lambda}}{\nu} 
  + T_{02} - \frac{T_{12}}{x-1}  \Bigr) \,
  \CF(x)  = 0 \,.
\end{align} 
For $N=1$, it can be derived from (\ref{ta}) that 
\begin{equation}\label{TFN1}
 \Bigl(  T_{01} + T_{02} + T_{12}
  + \fr{2}{\nu} h_{\la} \Bigr) 
 \big\langle \phi_{\la}(w)%
 \big\rangle^{\Lambda^*\!,\infty}_{\Lambda,0}  =0 \,.
\end{equation}
Therefore, equations (\ref{kz3a}) and~(\ref{kz3c}) are 
equivalent and it suffices to consider only the first of them. 

For $N=1$, the martingale condition (\ref{dPH2}) reads:
\begin{equation}\label{ABN}
 \CM(x)\,  \CF(x) =0 \,, \qquad
 \CM(x) = A_1  + x^2 A_2 + x B_{12} \,,
\end{equation}
where $A_1$, $A_2$, and $B_{12}$ are given by (\ref{Aj})
and (\ref{Bj}).
We will analyse condition (\ref{ABN}) for the weights 
$\Lambda$ allowed by the necessary conditions 
(see the cases 0)--2b) in Section~2), and for the weights
$\lambda$ such that the submodule $W^{\mathfrak{g}}$ be 
two--dimensional.
Recall (see Section~2) that  $k>1$ implies $\mathfrak{g}=\mathfrak{su}(2)$,
and this case  was  analysed in detail in Section~3.
For this reason, we will restrict our consideration to the case of $k=1$. 
Although $W^{\mathfrak{g}}$ is assumed to be two--dimensional,
the space of conformal blocks for $k=1$ 
is one--dimensional due to the fusion rules
(for a recent account see~\cite{RV}).

\subsection{$\mathfrak{g}=\mathfrak{su(n)}$, 
  $\Lambda$ $n$--dimensional representation}

Let $\mathfrak{g}=\mathfrak{su}(n)$, $n \geq 2$, 
and let $\Lambda$ be the $n$--dimensional defining representation.
If $\lambda$ coincides with $\Lambda$ or $\Lambda^*$,
then $W^{\mathfrak{g}}$ is two--dimensional.
For definiteness, we take $\lambda=\Lambda^*$.

If a pair of vectors, $v_1$ and~$v_2$, forms a basis of 
$W^{\mathfrak{g}}$, then a solution to the KZ equation
(\ref{kz3a})  has the form
$\CF(x) = F_1(x) v_1 + F_2(x) v_2$, where 
$F_1(x)$ and $F_2(x)$ are scalar functions. 
We take $v_1= \varepsilon_{12} \varepsilon_{03}$,  where
$\varepsilon$ is the normalised basis vector
of the trivial representation appearing in the 
decomposition of $V_\Lambda \ot V_{\Lambda^*}$. We choose such
$v_2$ that the pair $v_1$, $v_2$ forms an orthonormal 
basis. Recall that $T_{ij}=T_{ij}^*$. Therefore, 
$T_{12}^{\mathfrak{g}}$ is represented by a diagonal
matrix, and $T_{01}^{\mathfrak{g}}$, $T_{02}^{\mathfrak{g}}$
are represented by symmetric matrices.
Their eigenvalues can be found using the following formula:
\begin{equation}\label{TCC}
T_{ij}=\fr{1}{2} \bigl( C_{ij} - C_i -C_j \bigr) .
\end{equation}
Since $V_\Lambda\ot V_\Lambda=
V(2,0,{\ldots}) \oplus V(0,1,0,{\ldots})$,
the eigenvalues of $T_{02}^{\mathfrak{g}}$
are $\bigl(\frac{n-1}{n},\frac{-n-1}{n}\bigr)$.
Since $V_\Lambda\ot V_{\Lambda^*}=V(0,{\ldots},0) \oplus
V(1,0,{\ldots},0,1)$, the eigenvalues of 
$T_{01}^{\mathfrak{g}}$ and $T_{12}^{\mathfrak{g}}$ 
are $\bigl(\frac{1-n^2}{n},\frac{1}{n}\bigr)$.
This, along with relation (\ref{TFN1}), determines
entries of the sought matrices uniquely (up to the sign 
of the off--diagonal entries of 
$T_{01}^{\mathfrak{g}}$, $T_{02}^{\mathfrak{g}}$
which can be reverted by changing $v_2 \to -v_2$):
\begin{equation}\label{TTT} 
 T_{01} = -\fr{1}{n} \left(\!\begin{smallmatrix} 
  0 & \sqrt{n^2-1} \\[1mm]
 \sqrt{n^2-1} & n^2-2 \end{smallmatrix}\right)  ,\qquad 
 T_{02} = \fr{1}{n} \left(\!\begin{smallmatrix} 
  0 & \sqrt{n^2-1} \\[1mm] 
 \sqrt{n^2-1} & -2 \end{smallmatrix}\right) , \qquad
 T_{12} = 
\fr{1}{n} \left(\!\begin{smallmatrix} 1-n^2  & 0 \\[1mm] 
 0 & 1 \end{smallmatrix}\!\right).
\end{equation}
Here we use the identification
$v_1 \sim \left(\!\begin{smallmatrix} 1 \\ 0\end{smallmatrix}\!\right)$,
$v_2 \sim \left(\!\begin{smallmatrix} 0 \\ 1\end{smallmatrix}\!\right)$. 

\subsubsection{ $n=2$}

Substituting (\ref{TTT}) for $n=2$ in (\ref{ABN}) and  
setting $\tau=2 -\varkappa/2$ according to (\ref{taukap}), 
we obtain the following matrix  $\CM(x)$ for $N=1$, $k=1$:
\begin{equation}\label{MMM} 
 \CM(x) = (3-\varkappa) 
 \left(\!\begin{smallmatrix} 2 (x-1)^2  &
 \frac{2}{\sqrt{3}} (x^2-1) \\[1mm] 
 \frac{2}{\sqrt{3}} (x^2-1) & 
 \frac{2}{3} (x+1)^2 \end{smallmatrix}\!\right) .
\end{equation}
For $\varkappa \neq  3$ and generic $x$, the rank of $\CM(x)$ is one. 
The eigenvector corresponding to the zero eigenvalue is 
\begin{equation}\label{F02}
\CF_{0}(x) =  (x+1) v_1 + \sqrt{3}\, (1-x) v_2  \, .
\end{equation}
Using again (\ref{TTT}), it is straightforward to check that
\begin{equation}\label{F022}
\CF(x) =   \bigl(x(1-x)\bigr)^{-\frac{1}{2}} \CF_0(x) 
\end{equation}
satisfies the KZ equation~(\ref{kz3a}). Note that
$-\frac{1}{2} = -2 h_\lambda$ is consistent with~(\ref{bound2}).
Thus, we conclude that, in the $\mathfrak{su}(2)_1$ case,
the boundary one--point correlator is an SLE$_\varkappa$
martingale for any~$\varkappa$.

\subsubsection{ $n>2$ }

Substituting (\ref{TTT}) in (\ref{ABN}) and setting
$\varkappa=2$, $\tau=2/n$ according to (\ref{oth6}),
we obtain the following matrix  $\CM(x)$ for $N=1$, $k=1$:
\begin{equation}\label{MM} 
 \CM(x) = \fr{2(n^2+n-2)}{n^2} 
 \left(\!\begin{smallmatrix} 
 (x-1)^2 & \frac{(x-1)(nx-x+1)}{\sqrt{n^2-1}} \\
 \frac{(x-1)(nx-x+1)}{\sqrt{n^2-1}} & \frac{(nx-x+1)^2}{n^2-1}
\end{smallmatrix}\!\right) .
\end{equation}
For $n>1$ and generic $x$, the rank of $\CM(x)$ is one. 
The eigenvector corresponding to the zero eigenvalue is 
\begin{equation}\label{F0n}
\CF_{0}(x) =   (nx-x+1) v_1 + \sqrt{n^2-1}\, (1-x) v_2 \,.
\end{equation}
Using (\ref{TTT}), it is straightforward to check that
\begin{equation}\label{F0nn}
\CF(x) =   \bigl(x(1-x)\bigr)^{\frac{1}{n}-1} \CF_0(x) 
\end{equation}
satisfies the KZ equation~(\ref{kz3a}). Note that
$\frac{1}{n}-1 = -2 h_\lambda$ is consistent with~(\ref{bound2}).
We conclude that
the boundary one--point correlator is an SLE$_2$
martingale (recall that the space of conformal
blocks  is one--dimensional, cf.~Theorem 4.7 in~\cite{RV}).

\subsection{$\mathfrak{g}=\mathfrak{su(n)}$, 
  $\Lambda$ self--adjoint representation}

Let $\mathfrak{g}=\mathfrak{su}(n)$, $n \geq 2$ is even, 
and let $\Lambda=\Lambda^* =\omega_{n/2}$ ($\omega_{i}$
denotes the $i$--th fundamental weight, 
$\omega_{i}^*=\omega_{n-i}$). 
If $\lambda$ or $\lambda^*$ is equal to $\omega_1$ 
(the highest weight of the defining $n$--dimensional representation), 
then $W^{\mathfrak{g}}$ is two--dimensional.
Indeed, we have
\begin{equation}\label{om}
\begin{array}{lll}
  V\big(\omega_{\frac{n}{2}}\big) \ot V\big(\omega_{1}\big) & =  &
 V\big(\omega_{\frac{n}{2}+1}\big) \oplus 
 V\big(\omega_{1} + \omega_{\frac{n}{2}}\big) , \\[1mm]
  V\big(\omega_{n-1}\big) \ot V\big(\omega_{\frac{n}{2}}\big) & = &
 V\big(\omega_{\frac{n}{2}-1}\big) \oplus 
 V\big(\omega_{\frac{n}{2}} + \omega_{n-1} \big) ,
\end{array}
\end{equation}
where $V(\omega)$ is the irreducible representation
of highest weight $\omega$.
In the decomposition of the tensor product of the l.h.s., 
the trivial representation appears once in the product
of the first terms and once in the product of the second
terms on the r.h.s.
For definiteness, we take $\lambda$ to be the fundamental 
representation. 

As a basis of $W^{\mathfrak{g}}$ we take
$v_1= \varepsilon_{12} \varepsilon'_{03}$ 
($\varepsilon$ and $\varepsilon'$ are the normalised basis 
vectors of the trivial representation appearing in the 
decomposition of $V_\lambda \ot V_{\lambda^*}$ and
$V_\Lambda \ot V_\Lambda$, respectively) and such $v_2$
that the basis be orthonormal. Then, 
$T_{12}^{\mathfrak{g}}$ is represented by a diagonal
matrix, and $T_{01}^{\mathfrak{g}}$, $T_{02}^{\mathfrak{g}}$
are represented by symmetric matrices.
Their eigenvalues are found from (\ref{TCC}) and 
(\ref{om}), and they are equal to 
$\bigl(\frac{1-n^2}{n},\frac{1}{n}\bigr)$ for
$T_{12}^{\mathfrak{g}}$, and 
$\bigl(\frac{-n-1}{2},\frac{1}{2}\bigr)$ for
$T_{01}^{\mathfrak{g}}$ and $T_{02}^{\mathfrak{g}}$.
This, along with relation (\ref{TFN1}), determines
entries of the sought matrices uniquely (up to the sign 
of the off--diagonal entries of 
$T_{01}^{\mathfrak{g}}$, $T_{02}^{\mathfrak{g}}$
which can be reverted by changing $v_2 \to -v_2$):
\begin{equation}\label{TTTT} 
 T_{01} = -\fr{1}{2} \left(\!\begin{smallmatrix} 
  0 & \sqrt{n-1} \\[1mm]
 \sqrt{n-1} & n \end{smallmatrix}\right)  ,\qquad 
 T_{02} = \fr{1}{2} \left(\!\begin{smallmatrix} 
  0 & \sqrt{n-1} \\[1mm] 
 \sqrt{n-1} & -n \end{smallmatrix}\right) , \qquad
 T_{12} = 
\fr{1}{n} \left(\!\begin{smallmatrix} 1-n^2  & 0 \\[1mm] 
 0 & 1 \end{smallmatrix}\!\right).
\end{equation}
Substituting (\ref{TTTT}) in (\ref{ABN}) and setting 
$\varkappa = 8/(n+2)$, $\tau = 4/(n+2)$
according to (\ref{oth7}),
we obtain the following matrix  $\CM(x)$ for $N=1$, $k=1$:
\begin{equation}\label{MMMM} 
 \CM(x) = \fr{2n^2}{n+2}
 \left(\!\begin{smallmatrix} 
 (x-1)^2 & \frac{x^2-1}{\sqrt{n+1}} \\ 
 \frac{x^2-1}{\sqrt{n+1}} & \frac{(x+1)^2}{n+1} 
\end{smallmatrix}\!\right) .
\end{equation}
For generic $x$, the rank of $\CM(x)$ is one. 
The eigenvector corresponding to the zero eigenvalue is 
\begin{equation}\label{F0nN}
\CF_{0}(x) =   (x+1) v_1 + \sqrt{n+1}\, (1-x) v_2 \,.
\end{equation}
Using (\ref{TTTT}), it is straightforward to check that
\begin{equation}\label{F0nnNN}
\CF(x) =   (1-x)^{\frac{1}{n}-1} x^{-\frac{1}{2}} \CF_0(x) 
\end{equation}
satisfies the KZ equation~(\ref{kz3a}). Note that
$\frac{1}{n}-1 = -2 h_\lambda$ is consistent with~(\ref{bound2}).
We conclude that
the boundary one--point correlator is an SLE$_\varkappa$
martingale for $\varkappa = 8/(n+2)$.

\section{SLE$_{\varkappa,\rho}$ version}

The SLE$_{\varkappa,\rho}$ process was introduced in \cite{LSW}
as a generalization of the SLE process. 
More specifically, consider a random curve $\gamma_t$ which
starts at the origin of the upper half plane~$\HH$,
and chose a point $r \in \mathbb R$. In the SLE$_{\varkappa,\rho}$
process, the dynamics of the tip $z_t$ of the curve 
is given by the law $z_t=g^{-1}_t(x_t)$, where
$g_t(z)$ is the uniformizing conformal map which maps 
the slit domain $\HH\slash{\gamma_t}$ 
back to~$\HH$ and which satisfies the following 
stochastic differential equation:
\begin{equation}\label{dGxt}
  dg_t(z) = \frac{2dt}{g_t(z) - x_t } \,, \qquad g_0(z) = z \,,
\end{equation}
where $x_t = g_t(z_t) \in \mathbb R$ in turn satisfies
\begin{equation}\label{dxt}
  dx_t = \sqrt{\varkappa}\, d\xi_t + \frac{\rho\, dt}{x_t- g_t(r)} \,,
 \qquad x_0 =0 \,.
\end{equation}
Here $\xi_t$ is the normalized Brownian process on $\mathbb R$
starting at the origin, $\varkappa$ is the diffusion
coefficient, and $\rho \in \mathbb R$ is the drift parameter.

This setting can be used to study martingale properties of
boundary correlation functions in the case when the second
boundary changing operator is inserted 
at the finite distance from the origin (instead of infinity).
For the $U(1)$ CFT, this approach was used in~\cite{Ca3}.
In the WZW case, we consider the boundary correlator
\begin{equation}\label{bound3}
\begin{aligned}
 \bcr {}& \equiv  \frac{
 \langle \phi_{\Lambda}(z_t)
  \phi_{\la_1}(z_1) \ldots \phi_{\la_N}(z_N) \,
  \phi_{\la_1^*}(\bar{z}_1) \ldots
    \phi_{\la_N^*}(\bar{z}_N) \phi_{\Lambda^*}(r) \rangle^\mathfrak{g}
 }{ \langle \phi_{\Lambda}(z_t) \phi_{\Lambda^*}(r) 
 \rangle^\mathfrak{g} } \,.
\end{aligned}
\end{equation}
Conformal covariance implies that 
$\langle\phi_{\Lambda}(z_t) \phi_{\Lambda^*}(r) \rangle^\mathfrak{g}
 = \text{const}\ |z_t -r|^{-2h_\Lambda}$.

It is convenient to introduce a conformal map
$w_t(z) = g_t(z) -x_t$, so that the dynamics of the tip of the 
trace is given by $z_t=w^{-1}_t(0)$.
The map $w_t(z)$ maps the initial configuration of fields on
the slit domain $\HH\slash\gamma_t$ into a configuration on the upper 
half plane~$\HH$. The boundary condition changing operators 
$\phi_\Lambda$ and $\phi_{\Lambda^*}$ are now inserted at the points
$w=0$ and $y \equiv w_t(r)=g_t(r) - x_t$, and the bulk primary fields 
$\phi_{\lambda_i}$ are positioned at the points $w_i \equiv w_t(z_i)$.
It follows from (\ref{dGxt})--(\ref{dxt}) that
\begin{equation}\label{dwy}
 dw_i = \bigl( \frac{2}{w_i} + \frac{\rho}{y} \bigr) \, dt
 - \sqrt{\varkappa} \, d\xi_t \,, 
 \qquad  w_i\bigm|_{t=0} = z_i \,, 
\end{equation}
where $i=1,{\ldots},2N\,{+}\,1$ with understanding that
$w_{i+N}=\bar{w}_i$ and $w_{2N+1} \equiv y$.
Note that as $t$ increases the boundary field $\phi_{\Lambda^*}$
moves along the boundary.

Using the conformal map $w_t(z)$, we rewrite correlator 
(\ref{bound3}) in the new coordinates: 
\begin{align}
\label{bc3}
  \bcr =
 \biggl( \prod_{i=1}^{2N}
 \Bigl( \frac{\partial w_i}{\partial z_i} \Bigr)^{h_{i}}
 \!\biggr) \, y^{2h_\Lambda} \bcy \,,
\end{align}
where
\begin{equation}\label{bound4}
 \bcy \equiv  
 \langle \phi_{\Lambda}(0)
  \phi_{\la_1}(w_1) \ldots \ldots
    \phi_{\la_N^*}(w_{2N}) \phi_{\Lambda^*}(y) \rangle^\mathfrak{g} .
\end{equation}

When $t$ is increased by~$dt$, the fields at $w_i$, 
$i=1,{\ldots},2N\,{+}\,1$ are changed as follows
\begin{equation}\label{dff}
 d\phi_{\la_i}(w_i) = {\cal G}_i \phi_{\la_i}(w_i) \,, 
\end{equation}
where
\begin{equation}\label{dPP}
 {\cal G}_i  =
 \Bigl( \bigl( \frac{2}{w_i} + \frac{\rho}{y} \bigr) dt 
 - \sqrt{\varkappa} \, d\xi_t \Bigr) \partial_{w_i} 
 + \sqrt{\tau} \bigl(\frac{1}{w_i} - \frac{1}{y} \bigr)
 \sum_{a=1}^{\text{dim}\,{\mathfrak g}} 
 \Bigl(  d\theta^a \, t^a_i \Bigr)   \,.
\end{equation}
Here the first term is due to (\ref{dwy}) while the coordinate 
dependent coefficient in the second term is obtained from the 
analogous coefficient in (\ref{dP}) by the inverse to
the M\"obius map $\tilde{w}=y w/(w+y)$ (which maps the infinity 
to $y$ while preserving the origin).

A computation similar to (\ref{boundGG1}) shows that
\begin{align}
\nonumber
\EE\bigl[\frac{d}{dt} \bc \bigr] =
\biggl( y^{2h_\Lambda} \prod_{i=1}^{2N} \Bigl(  
 \frac{\partial w_i}{\partial z_i} \Bigr)^{h_{i}} \!\biggr)\,
 \tilde{\Theta} \, \bcy \,,
\end{align} 
where
\begin{align}
\nonumber
\tilde{\Theta} &=
\biggl( \frac{h_\Lambda}{y^2} 
   \bigl(6+ 2\rho +\varkappa(2h_\Lambda-1)\bigr)
 + \sum_{i=1}^{2N+1} \Bigl(
 \frac{2}{w_i} \, \partial_{w_i} - \frac{2h_i}{w_i^2} 
 + \frac{\varkappa}{2}  \partial_{w_i} \partial_{w_j} \Bigl) 
 \\
\nonumber
{}& 
 + (\rho + 2\varkappa h_\Lambda) \frac{1}{y} 
 \sum_{i=1}^{2N+1} \partial_{w_i} 
 + \frac{\tau}{2} \sum_{i,j=1}^{2N+1} \Bigl(
 \frac{ T_{ij} }{w_i\,w_j} 
 - \frac{2 T_{ij}}{y\, w_i} 
 + \frac{T_{ij}}{y^2} \Bigr)
 \biggr) .
\end{align}
Thus, the correlator (\ref{bound3}) is an SLE$_{\varkappa,\rho}$
martingale if and only if  $\bcy$ is annihilated by $\tilde{\Theta}$.
Using (\ref{CL}) and the Sugawara relations, 
$L_0=\frac{\nu}{2} \sum_a J^a_0 J^a_0$,
$L_{-1}= \nu \sum_a J^a_{-1} J^a_0$, we can rewrite this
condition as follows
\begin{align}
\nonumber
{}&  \Bigl( -2L_{-2} + \frac{\varkappa}{2} L_{-1}^2 
 + \frac{\tau}{2} \sum_{a=1}^{\dim \mathfrak g} J^a_{-1} J^a_{-1}
 - (\rho + 2 \varkappa h_\Lambda + \frac{\tau}{\nu})
 \frac{1}{y} L_{-1}  \\
\label{elj}
{}& 
 + \frac{h_\Lambda}{y^2} 
 \bigl(6+2\rho +\varkappa(2h_\Lambda-1) + \frac{\tau}{\nu} \bigr)
 \Bigl) \bcy =0 \,.
\end{align} 
Here $L$'s and $J$'s act on $\phi_\Lambda(0)$.

Note that the terms involving $\rho$ and $y$ are of level $0$ and $-1$.
Therefore, applying level two operators $J^b_2$, $L_1 J^b_1$, and $L_2$ 
to (\ref{elj})  yields the same relations 
(\ref{taukap}), (\ref{jp3}), and (\ref{jp4}).
A combination of the first two of them leads to the following
constraint:
\begin{equation}\label{kapparhotau}
2 \varkappa h_\Lambda + \varkappa + \fr{\tau}{\nu} =6 \,.
\end{equation}
Furthermore, by applying $L_1$ to (\ref{elj}) we obtain
\begin{equation}\label{yrho}
 \fr{2}{y} h_\Lambda (\rho + 2 \varkappa h_\Lambda
   + \fr{\tau}{\nu} ) \, \phi_{\Lambda} =
 \bigl( \varkappa (2 h_\Lambda +1)
   + \fr{\tau}{\nu} -6 \bigr) \, L_{-1} \, \phi_{\Lambda} .
\end{equation}
Assuming that $L_{-1}\, \phi_\Lambda \neq 0$, we conclude from 
(\ref{kapparhotau}) and (\ref{yrho}) that $\rho$ is uniquely
determined:
\begin{equation}\label{kapparho}
 \rho = \varkappa - 6 \,.
\end{equation}

It is well known \cite{SW} that, up to a time change, 
the SLE${}_{\varkappa,\varkappa-6}$ process is an image of 
the SLE${}_\varkappa$ process under the M\"obius map  preserving 
zero and mapping $\infty$ to a finite point $y$.
As a consequence, SLE${}_{\varkappa,\varkappa-6}$ describes 
a random curve which
starts at the origin and aims at the point $y$ on the real axis.
Choose a M\"obius map preserving the singularity at $0$ (that is, 
$d\tilde{w}/dw|_{w=0}=1$) and, hence, preserving the 
parametrization of the 
Loewner chain.  Then, the coefficient $\frac{1}{w_i}-\frac{1}{y}$ 
in front of the second term of (\ref{dPP}) is exactly the 
push-forward of its counterpart in (\ref{dP}), and 
the whole SLE${}_{\varkappa,\rho}$ picture is a M\"obius image 
of the SLE${}_{\varkappa}$ one.

Taking constraints (\ref{kapparhotau}) and (\ref{kapparho}) into
account, we see that the martingale condition (\ref{elj})
reduces to
\begin{equation}\label{mc2}
 \Bigl( -2L_{-2} + \frac{\varkappa}{2} L_{-1}^2 
  + \frac{\tau}{2} \sum_{a=1}^{\dim \mathfrak g} J^a_{-1} J^a_{-1}
 \Bigr) \bcy =0 \,.
\end{equation}
Thus, a sufficient condition for the boundary correlator (\ref{bound3})
to be a covariant $\text{SLE}_{\varkappa,\varkappa -6}$ martingale
is the same as in the $\text{SLE}_{\varkappa}$ case, i.e. that the
operator $\psi$ given by (\ref{psi}) be a level two null vector.
Therefore, all the results of Section~2 apply here as well.

A counterpart of the necessary condition~(\ref{dPH2}) can be 
obtained as follows. The correlator $\bcy$ satisfies the KZ equation 
(\ref{kz2}), where $2N$ is replaced by $2N\,{+}\,1$ 
(recall that $w_{2N+1} \equiv y$). Using the corresponding versions 
of eqs. (\ref{kzNN})--(\ref{kzNNN}) along with the following relations
\begin{equation}\label{T1} 
 \bigl( T_{0i} + T_{i,{2N+1}} + 
   \sum_{j=1}^{2N} T_{ij} \bigr)^{\mathfrak{g}} 
   = 0 \,, \qquad
 \bigl( \sum_{1 \leq i<j}^{2N} T_{ij} + 
 \fr{1}{\nu} \sum_{i=1}^{2N} h_i 
   - \fr{2}{\nu} h_\Lambda - T_{0,{2N+1}} \bigr)^{\mathfrak{g}} =0
\end{equation}
which are consequences of the $\mathfrak{g}$--invariance, cf.~(\ref{ta}),
we repeat the computations of Section~3. As a result, we find
that the condition $\tilde{\Theta} \bcy=0$ is equivalent to 
the condition that $\bcy$ belongs to the kernel of a certain matrix:
\begin{equation}\label{dPH2t} 
  \tilde{\CM}\bigl(\{w_{i}\}\bigr) \, \bcy =0 \,,  
\end{equation}
where 
 \begin{equation} \label{tMw}
  \tilde{\CM} \bigl(\{w_{i}\}\bigr) =
  \sum_{i=1}^{2N+1} \frac{A_i}{w_{i}^2} +
  \sum_{1\leq i<j}^{2N+1} \frac{B_{ij}}{w_{i} w_{j}} ,
\end{equation}
with $A_i$ and $B_{ij}$ given by the same formulae (\ref{Aj})
and (\ref{Bj}), respectively. The difference with the
SLE$_{\varkappa}$ case is that $\tilde{\CM}\bigl(\{w_{i}\}\bigr)$ 
contains more terms. Eqs. (\ref{AB0}) imply that 
$\tilde{\CM} \bigl(\{w_{i}\}\bigr)$ vanishes identically
in the case $\mathfrak{g}=\mathfrak{su}(2)$, $\Lambda$
being the fundamental representation.

\par\vspace*{3mm}\noindent
{\small
{\bf Acknowledgements.}
\noindent 
 The authors thank Stanislav Smirnov for stimulating discussions,
 and Paul Wiegmann and J\"org Teschner for useful comments. 
\noindent
 This research was supported in part by the Swiss National 
 Science Foundation (grants 200020--126817 and 200020--129609).
 The work of A.B. was supported in part by the Russian Foundation
 for Fundamental Research (grants 07--02--92166, 
 09--01--12150, 09--01--93108). The work of K.I. was partially 
supported by ERC AG CONFRA.
The research of A.B. and K.I.  received support of the 
Chebyshev Laboratory (Faculty of Mathematics
and Mechanics, St.~Petersburg State University) under the grant
11.G34.31.2006 of the Government of the Russian Federation.
}

\def\baselinestretch{1}

\end{document}